\documentclass[openright,11pt]{article}

\pdfoutput=1
\usepackage{enumerate}
\usepackage{comment}
\usepackage{amsmath}
\usepackage{amssymb}
\usepackage{color}
\usepackage{pdfpages}
\usepackage[dvipsnames]{xcolor}
\usepackage{tikz}
\usepackage{amsthm}
\usepackage{graphicx}
\usepackage{caption}
\usepackage{subcaption}
\usepackage{bbold}
\usepackage{empheq}
\usepackage{pdflscape}
\usepackage{afterpage}
\usepackage{array, diagbox}
\usepackage{adjustbox}
\usepackage{sectsty}
\sectionfont{\centering}
\usepackage[T1]{fontenc}

\usepackage[%
  autocite    = superscript,
  backend     = bibtex,
  sortcites   = true,
  style       = numeric,
  ]{biblatex}

\newcommand{\Hh}{\text{H}}

\newcommand{\eqn}{equation }
\newcommand{\eqns}{equations }

\begin{document}

\includepdf[pages={1,2,3,4,5,6,7,8,9,10,11,12,13}]{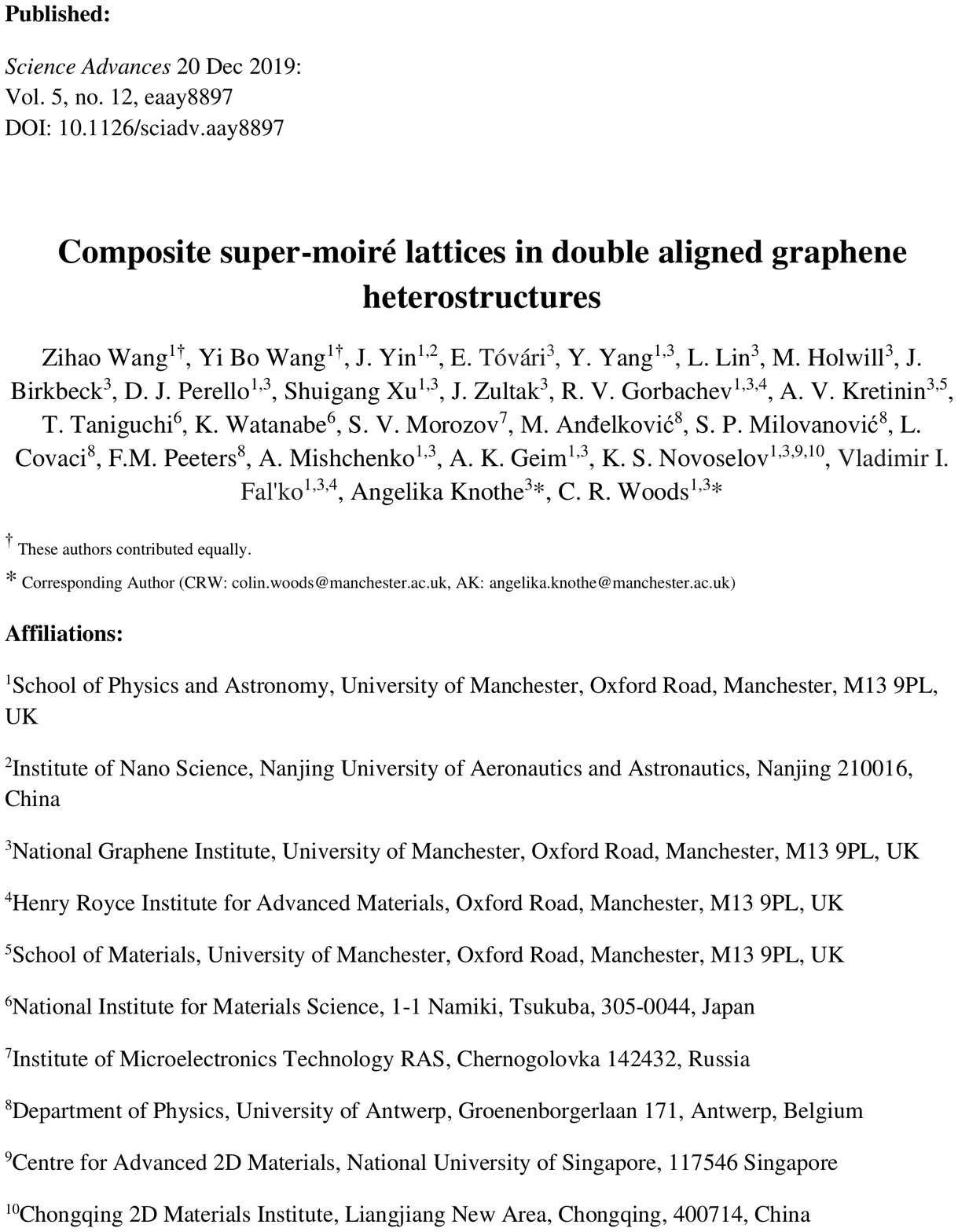}

\section*{Composite super-moir\'e lattices in double aligned graphene heterostructures}

\begin{center}
\textbf{Authors: }  Zihao Wang$^{1\dagger}$, Yi Bo Wang$^{1\dagger}$, J. Yin$^{1,2}$, E. T\'ov\'ari$^{3}$, Y. Yang$^{1,3}$, L. Lin$^{3}$, M. Holwill$^{3}$, J. Birkbeck$^{3}$, D. J. Perello$^{1,3}$, Shuigang Xu$^{1,3}$, J. Zultak$^{3}$, R. V. Gorbachev$^{1,3,4}$, A. V. Kretinin$^{3,5}$, T. Taniguchi$^{6}$, K. Watanabe$^{6}$, S. V. Morozov$^{7}$, M. An\dj{}elkovi\'c$^{8}$, S.P. Milovanovi\'c$^{8}$, L. Covaci$^{8}$, F.M. Peeters$^{8}$, A. Mishchenko$^{1,3}$, A. K. Geim$^{1,3}$, K. S. Novoselov$^{1,3,9,10}$, Vladimir I. Fal'ko$^{1,3,4}$, Angelika Knothe$^{3*}$, C. R. Woods$^{1,3*}$\\
\vskip5pt
{\small $*$ Correspondence to: CRW `colin.woods@manchester.ac.uk', AK `angelika.knothe@manchester.ac.uk'}
\end{center}

\begin{center} \subsection*{
Supplementary Information}
\end{center}

\subsubsection*{Contents:}

\hspace*{35pt}Details of the super-moir\'e superlattice perturbation theory\\

\subsubsection*{Details of the super-moir\'e superlattice perturbation theory}

We study the long-range periodic "super"-moire pattern which appears due to beatings between the two moires at the top and bottom interfaces. The six reciprocal lattice vectors of the top (bottom) moire pattern are given by $\mathbf{b}^{\alpha}_m=\mathbf{G}_m-\mathbf{g}^{\alpha}_m$ ($\mathbf{b}^{\beta}_m=\mathbf{G}_m-\mathbf{g}^{\beta}_m$) for $m=0,\dots,5$, where $\mathbf{G}_m$ denote the  reciprocal lattice vectors of graphene, and $\mathbf{g}^{\alpha}_m$ ($\mathbf{g}^{\beta}_m$) are the reciprocal lattice vectors of the top (bottom) hBN. From these, we construct the combinations $\mathbf{d}_{m,k}= \mathbf{b}^{\alpha}_{m}-\mathbf{b}^{\beta}_k$. For twist different angles $\theta^{\alpha}$ ($\theta^{\beta}$) of the top (bottom) hBN layer, these become very small or vanish completely and hence constitute the shortest reciprocal lattice vectors of the "super"-moire pattern. These cases are studied below.

 \subsubsection*{Derivation of the Hamiltonians of shortest period}

The low-energy contribution (for the shortest effective Bragg vectors $\mathbf{d}_{m,k}=\mathbf{b^{\alpha}}_{m}-\mathbf{b}_{k}^{\beta}$) of the superlattice Hamiltonian which originates from interference reads in second order perturbation theory
\begin{align}
\nonumber\Hh_{n,m}^{int} = \delta\Hh^{(2)}_{\alpha\beta} +  \delta\Hh^{(2)}_{\beta\alpha},
\end{align}
with \\
\resizebox{1.05\linewidth}{!}{
  \begin{minipage}{\linewidth}
\begin{align}
\nonumber& \delta\Hh^{(2)}_{\substack{\alpha\beta\\ (\beta\alpha)}} = \sum_{m,k}\frac{v}{v^2 b^2}\Biggl( \bigg[ U_0 U_1 (-1)^{k} \frac{1}{a} \boldsymbol{b}_{k}^{\beta(\alpha)}.\boldsymbol{a}_{k}+U_0 U_1 (-1)^{m } \frac{1}{a} \boldsymbol{b}_{k}^{\beta(\alpha)}.\boldsymbol{a}_{m }\\
 \nonumber &+ U_1U_3 (-1)^{m+k}\frac{1}{a}(\boldsymbol{\ell}_z\times\boldsymbol{b}_{k}^{\beta(\alpha)}).(\boldsymbol{a}_{k}+\boldsymbol{a}_{m}) \bigg]e^{\mp i\frac{\mathbf{R}}{2}(\boldsymbol{b}_{k}^{\beta}+\boldsymbol{b}_{m}^{\alpha})} \; e^{i(\boldsymbol{b}_{k}^{\beta}-\boldsymbol{b}_{m}^{\alpha}).\mathbf{r}}\\
  \nonumber &+ i \frac{1}{a}\bigg[ (-1)^{k} U_1 U_0 (\boldsymbol{\ell}_z\times\boldsymbol{b}_{k}^{\beta(\alpha)}).\boldsymbol{a}_{k} - (-1)^{m } U_1 U_0 (\boldsymbol{\ell}_z\times\boldsymbol{b}_{k}^{\beta(\alpha)}).\boldsymbol{a}_{m }\\
 \nonumber &-   U_1 U_3 (-1)^{m+k} \boldsymbol{b}_{k}^{\beta(\alpha)}.(\boldsymbol{a}_{k}-\boldsymbol{a}_{m})  \bigg] \sigma_3 \, e^{\mp i\frac{\mathbf{R}}{2}(\boldsymbol{b}_{k}^{\beta}+\boldsymbol{b}_{m}^{\alpha})} \;e^{i(\boldsymbol{b}_{k}^{\beta}-\boldsymbol{b}_{m}^{\alpha}).\mathbf{r}}\\
  \nonumber &+ \bigg[  i (-1)^{m+k} U_1^2 \frac{1}{a^2} \big[ (\boldsymbol{a}_{m^{ }}.\boldsymbol{b}_{k}^{\beta(\alpha)}).(\boldsymbol{a}_{k}.\boldsymbol{\sigma}) + [\boldsymbol{a}_{m}.(\boldsymbol{\ell}_z\times\boldsymbol{b}_{k}^{\beta(\alpha)})].[(\boldsymbol{\ell}_z\times\boldsymbol{a}_{k}).\boldsymbol{\sigma}] \big] \bigg]e^{\mp i\frac{\mathbf{R}}{2}(\boldsymbol{b}_{n^{\prime}}^{\beta}+\boldsymbol{b}_{m}^{\alpha})} \;e^{i(\boldsymbol{b}_{k}^{\beta}-\boldsymbol{b}_{m}^{\alpha}).\mathbf{r}}\\
\nonumber   &+\bigg[U_0^2 \boldsymbol{b}_{k}^{\beta(\alpha)} - (-1)^{m+k} U_3^2 \boldsymbol{b}_{k}^{\beta(\alpha)} +  U_0 U_3 \big( (-1)^{k} + (-1)^{m} \big) (\boldsymbol{\ell}_z\times\boldsymbol{b}_{n^{\prime}}^{\beta(\alpha)}) \bigg] \boldsymbol{\sigma} e^{\mp i\frac{\mathbf{R}}{2}(\boldsymbol{b}_{k}^{\beta}+\boldsymbol{b}_{m}^{\alpha})} \;e^{i(\boldsymbol{b}_{k}^{\beta}-\boldsymbol{b}_{m}^{\alpha}).\mathbf{r}} \Biggr),
\end{align}
\end{minipage}
}\begin{flushright} (S1)\end{flushright}
while the corresponding contribution due to strain caused by reconstruction is given by $(40)$ \\
\resizebox{1.05\linewidth}{!}{
  \begin{minipage}{\linewidth}
\begin{align}
\nonumber\Hh^{rec}=&\sum_{m,k} \Bigg(  i \Big[U_0^{\beta}\mathbf{G}_m \cdot \mathbf{u}_{\mathbf{b}_k^{\alpha}}-U_0^{\alpha}\mathbf{G}_m \cdot \mathbf{u}_{-\mathbf{b}_k^{\beta}}\Big] + i\sigma_3 \Big[ (-1)^m i U_3^{\beta}\mathbf{G}_m \cdot \mathbf{u}_{\mathbf{b}_k^{\alpha}} + (-1)^k i U_3^{\alpha  }\mathbf{G}_m \cdot \mathbf{u}_{-\mathbf{b}_k^{\beta}}\Big] \\
\nonumber&+   \Big[ (-1)^m i U_1^{\beta}  \frac{\mathbf{a}_m}{a}\boldsymbol{\sigma}\mathbf{G}_m \cdot \mathbf{u}_{\mathbf{b}_k^{\alpha}} - (-1)^k i U_1^{\alpha  } \frac{\mathbf{a}_k}{a}\boldsymbol{\sigma}\mathbf{G}_m \cdot \mathbf{u}_{-\mathbf{b}_k^{\beta}}\Big]  \Bigg)\; e^{i (\mathbf{b}_{m}^{ {\alpha }}+\mathbf{b}_{k}^{ {\beta}})\cdot\frac{\mathbf{R}}{2}  } \; e^{i \mathbf{d}_{mk}\cdot\mathbf{r}}.
\label{eqn:FulldnnHamSM}
\end{align}
\end{minipage}
}\begin{flushright} (S2)\end{flushright}
We consider the terms of the Hamiltonians above with $m = k$, under the assumption of very small angles (for which $U_i^{\beta}\approx U_i^{\alpha }=: U_i$):\\
\resizebox{1.05\linewidth}{!}{
  \begin{minipage}{\linewidth}
\begin{align*}
&\nonumber \delta\Hh^{int}_{mm} =  \delta\Hh^{(2)}_{\alpha\beta} +  \delta\Hh^{(2)}_{\beta\alpha} \\
\nonumber =& \sum_{m}\frac{v}{v^2 b^2}\Biggl( \bigg[ 2U_0U_1 (-1)^m \frac{1}{a}(\boldsymbol{b}_m^{\alpha}-\boldsymbol{b}_m^{\beta}).\boldsymbol{a}_m \\
\nonumber&+ 2U_1U_3 \frac{1}{a}\big[(\boldsymbol{\ell}_z\times \boldsymbol{b}_m^{\alpha}).\boldsymbol{a}_m+ (\boldsymbol{\ell}_z\times \boldsymbol{b}_m^{\beta}).\boldsymbol{a}_m \big]\bigg] e^{-i\frac{\mathbf{R}}{2}(\boldsymbol{b}_{m }^{\alpha}+\boldsymbol{b}_{m}^{\beta})} \; e^{i(\boldsymbol{b}_{m }^{\alpha}-\boldsymbol{b}_{m}^{\beta}).\mathbf{r}}\\
  \nonumber &+  \bigg[  i   U_1^2 \frac{1}{a^2} \Big[ [(\boldsymbol{a}_{m^{ }}.\boldsymbol{b}_m^{\alpha})-(\boldsymbol{a}_{m^{ }}.\boldsymbol{b}_m^{\beta})].(\boldsymbol{a}_{m }.\boldsymbol{\sigma}) \\
  \nonumber&+ \Big([\boldsymbol{a}_{m}.(\boldsymbol{\ell}_z\times\boldsymbol{b}_m^{\beta})]+[\boldsymbol{a}_{m}.(\boldsymbol{\ell}_z\times\boldsymbol{b}_m^{\alpha})]\Big).[(\boldsymbol{\ell}_z\times\boldsymbol{a}_{m }).\boldsymbol{\sigma}] \Big] \bigg]e^{-i\frac{\mathbf{R}}{2}(\boldsymbol{b}_{m }^{\alpha}+\boldsymbol{b}_{m}^{\beta})} \;e^{i(\boldsymbol{b}_{m }^{\alpha}-\boldsymbol{b}_{m}^{\beta}).\mathbf{r}}\\
 \nonumber    &+  \bigg[ U_0^2(\boldsymbol{b}_n^{\alpha}-\boldsymbol{b}_m^{\beta}) -U_3^2(\boldsymbol{b}_m^{\alpha}-\boldsymbol{b}_m^{\beta})     +2U_0U_3 (-1)^m [ (\boldsymbol{\ell}_z\times\boldsymbol{b}_m^{\beta}) + (\boldsymbol{\ell}_z\times\boldsymbol{b}_m^{\alpha})]\bigg] \boldsymbol{\sigma} e^{-i\frac{\mathbf{R}}{2}(\boldsymbol{b}_{m }^{\alpha}+\boldsymbol{b}_{m}^{\beta})} \;e^{i(\boldsymbol{b}_{m }^{\alpha}-\boldsymbol{b}_{m}^{\beta}).\mathbf{r}} \Biggr).
 \end{align*}\end{minipage}
}\begin{flushright} (S3)\end{flushright}

For the contribution due to strain, the cases in which the two hBN layers are either parallel, or antiparallel with respect to each other must be distinguished:\\

Parallel case: 
\begin{align*} 
\nonumber \Hh_{m,m}^{P}=&  - U_0 ( w_s^{\alpha} +  w_s^{\beta}) f^{(m,m)}_1 (\mathbf{r}) -U_3 ( w_s^{\alpha} -  w_s^{\beta}) f^{(m,m)}_2 (\mathbf{r}) \sigma_3 \\
\nonumber&+U_1 ( w_s^{\alpha} +  w_s^{\beta})\frac{\sqrt{3}a}{4\pi}\frac{1}{(\theta^{\beta}-\theta^{\prime})} \boldsymbol{\nabla}f^{(m,m)}_2 (\mathbf{r}) \boldsymbol{\sigma} \\
\nonumber&+U_0 ( w_{as}^{\alpha} -  w_{as}^{\beta}) f^{(m,m)}_2 (\mathbf{r})  -U_3 ( w_{as}^{\alpha} +  w_{as}^{\beta}) f^{(m,m)}_1 (\mathbf{r}) \sigma_3 \\
&+U_1 ( w_{as}^{\alpha} -  w_{as}^{\beta})\frac{\sqrt{3}a}{4\pi}\frac{1}{(\theta^{\beta}-\theta^{\alpha})} \boldsymbol{\nabla}f^{(m,m)}_1 (\mathbf{r}) \boldsymbol{\sigma},
\end{align*}\begin{flushright} (S4)\end{flushright}
Antiparallel case: 
\begin{align*} 
\nonumber \Hh_{m,m}^{AP}=&  - U_0 ( w_s^{\alpha} -  w_s^{\beta}) f^{(m,m)}_1 (\mathbf{r}) -U_3 ( w_s^{\alpha} +  w_s^{\beta}) f^{(m,m)}_2 (\mathbf{r}) \sigma_3 \\
\nonumber &+U_1 ( w_s^{\alpha} -  w_s^{\beta})\frac{\sqrt{3}a}{4\pi}\frac{1}{(\theta^{\beta}-\theta^{\alpha})} \boldsymbol{\nabla}f^{(m,m)}_2 (\mathbf{r}) \boldsymbol{\sigma} \\
&\nonumber+U_0 ( w_{as}^{\alpha} +  w_{as}^{\beta}) f^{(m,m)}_2 (\mathbf{r})  -U_3 ( w_{as}^{\alpha} -  w_{as}^{\beta}) f^{(m,m)}_1 (\mathbf{r}) \sigma_3 \\
&+U_1 ( w_{as}^{\alpha} +  w_{as}^{\beta})\frac{\sqrt{3}a}{4\pi}\frac{1}{(\theta^{\beta}-\theta^{\alpha})} \boldsymbol{\nabla}f^{(m,m)}_1 (\mathbf{r}) \boldsymbol{\sigma},
\end{align*}\begin{flushright} (S5)\end{flushright}

in terms of the functions $f^{(m,m)}_1(\mathbf{r})=\sum_m e^{i (\mathbf{b}_{m}^{\alpha}+\mathbf{b}_{m}^{  \beta})\cdot\frac{\mathbf{R}}{2}  }  e^{i\mathbf{d}_{m^{ },m}\cdot\mathbf{r}}$,  $f^{(m,m)}_2(\mathbf{r})=i\sum_m (-1)^m e^{i (\mathbf{b}_{m}^{\alpha}+\mathbf{b}_{m}^{ \beta})\cdot\frac{\mathbf{R}}{2}  } e^{i\mathbf{d}_{m^{ },m}\cdot\mathbf{r}}$ and  $\Re[\mathbf{G}_m\cdot \mathbf{u}_{\mathbf{b}^{\alpha}_m} ] = \Re[\mathbf{G}_m\cdot \mathbf{u}_{-\mathbf{b}^{\alpha}_m} ] = -(-1)^n w_{as}^{\alpha}$, $\Im[\mathbf{G}_m\cdot \mathbf{u}_{\mathbf{b}^{\alpha}_m}]= -\Im[\mathbf{G}_m\cdot \mathbf{u}_{-\mathbf{b}^{\alpha}_m}] =w_s^{\alpha}$.  \\

Under the assumption that for small and almost equal angles $w_i^{\alpha} \approx w_i^{\beta}\approx w_i$, and keeping in mind that the gradient terms in \eqns (S3), (S4), (S5) can be removed by a gauge transformation $(28)$ we arrive at the superlattice Hamiltonians $\Hh_{m,m}$ of \eqn (2) in the main text.

  \subsubsection*{All combinations of short effective moire Bragg vectors}
  
For all the possible shortest effective Bragg vectors $\mathbf{d}_{m,k}$ we list the functions that describe the  period and  density of the corresponding super-moire structures.\\

  $\bigstar\;\mathbf{d}_{m,m}=\mathbf{b}^{\alpha}_{m}-\mathbf{b}^{\beta}_{m}=\frac{4\pi}{\sqrt{3}\text{a}(1+\delta)}
\begin{pmatrix}
 \sin\theta^{\alpha}-\sin\theta^{\beta}\\
 \cos\theta^{\beta}-\cos\theta^{\alpha}
\end{pmatrix},
$

\begin{align}
\nonumber A_{m,m}=  \text{a} ({\delta}+1) \frac{1}{\sqrt{2-2 \cos (\theta^{\alpha}-\theta^{\beta})}},
\end{align}
\begin{align}
\nonumber n_{m,m}=  -\frac{16 [\cos (\theta^{\alpha}-\theta^{\beta})-1]}{\sqrt{3} \text{a}^2 ({\delta}+1)^2},
\end{align}

Divergence of $A_{m,m}$ and zero $n_{m,m}$ for all twist angles with $\theta^{\alpha}=\theta^{\beta}$.\\

  $\bigstar\;\mathbf{d}_{m,m+1}=\mathbf{b}^{\alpha}_{m}-\mathbf{b}^{\beta}_{m+1}=\frac{2\pi}{\text{a}(1+\delta)}
\begin{pmatrix}
 \frac{  3 {\delta}+2 \sqrt{3} \sin\theta^{\alpha}-\sqrt{3} \sin \theta^{\beta}-3 \cos\theta^{\beta}+3}{3} \\
 \frac{{\delta}-2 \cos \theta^{\alpha}-\sqrt{3} \sin \theta^{\beta}+\cos \theta^{\beta}+1 }{\sqrt{3}}  
\end{pmatrix},
 $
 \resizebox{1.05\linewidth}{!}{
  \begin{minipage}{\linewidth}
\begin{align*}
\nonumber& A_{m,m+1}\\
\nonumber=& \frac{2 \sqrt{3} \text{a} ({\delta}+1) }{ 3 \left({\delta}-2 \cos\theta^{\alpha}-\sqrt{3} \sin\theta^{\beta}+\cos\theta^{\beta}+1\right)^2+\left(3 {\delta}+2 \sqrt{3} \sin\theta^{\alpha}-\sqrt{3} \sin \theta^{\beta}-3 \cos\theta^{\beta}+3\right)^2 }  \\
\approx & \frac{  \text{a}({\delta}+1)}{ {\delta}^2+\sqrt{3} {\delta} (\theta^{\alpha}-\theta^{\beta})+(\theta^{\alpha})^2-\theta^{\alpha} \theta^{\beta}+(\theta^{\beta})^2} ,
\end{align*}\end{minipage}
}
\resizebox{1.05\linewidth}{!}{
  \begin{minipage}{\linewidth}
\begin{align*}
\nonumber &n_{m,m+1}\\
\nonumber&= \frac{ 2 \left(3 \left({\delta}-2 \cos \theta^{\alpha}-\sqrt{3} \sin \theta^{\beta}+\cos \theta^{\beta}+1\right)^2+\left(3 {\delta}+2 \sqrt{3} \sin\theta^{\alpha}-\sqrt{3} \sin \theta^{\beta}-3 \cos\theta^{\beta}+3\right)^2\right)}{3 \sqrt{3} \text{a}^2 ({\delta}+1)^2}\\
&\approx \frac{8 \left({\delta}^2-\sqrt{3} {\delta} (\theta^{\beta}-\theta^{\alpha})+(\theta^{\alpha})^2-\theta^{\alpha} \theta^{\beta}+(\theta^{\beta})^2\right)}{\sqrt{3} \text{a}^2 ({\delta}+1)^2},
\end{align*}\end{minipage}
}

Angles of the critical points (Divergence of $A$ and zero $n$):\\
\resizebox{1.05\linewidth}{!}{
  \begin{minipage}{\linewidth}
\begin{align*}
\theta^{\alpha}=-\theta^{\beta}= \tan ^{-1}\left[\frac{\sqrt{-({\delta}-1) ({\delta}+3)} {\delta}+\sqrt{-({\delta}-1) ({\delta}+3)}-\sqrt{3}}{{\delta} ({\delta}+2)-2} \right]   \approx-  \frac{{\delta}}{\sqrt{3}}.
\end{align*}\end{minipage}
}\\

 $\bigstar\;\mathbf{d}_{m,m+2}=\mathbf{b}^{\alpha}_{m}-\mathbf{b}^{\beta}_{m+2}=\frac{2\pi}{\text{a}(1+\delta)}
\begin{pmatrix}
 \frac{3 {\delta}+2 \sqrt{3} \sin \theta^{\alpha}+\sqrt{3} \sin \theta^{\beta}-3 \theta^{\beta}+3 }{3} \\
 \frac{ -3 {\delta}+2 \cos\theta^{\alpha}+\sqrt{3} \sin \theta^{\beta}+\cos \theta^{\beta}-3}{\sqrt{3}}  
\end{pmatrix},
$

\resizebox{1.05\linewidth}{!}{
  \begin{minipage}{\linewidth}
\begin{align*}
\nonumber &A_{m,m+2}=\\
\nonumber& \frac{2 \sqrt{3}  ({\delta}+1)\;\text{a}}{\sqrt{    3 \left(-3 {\delta}+2 \cos \theta^{\alpha}+\sqrt{3} \sin \theta^{\beta}+\cos \theta^{\beta}-3\right)^2+\left(3 {\delta}+2 \sqrt{3} \sin\theta^{\alpha}+\sqrt{3} \sin\theta^{\beta}-3 \cos\theta^{\beta}+3\right)^2} }\\
\approx &  \frac{\text{a}({\delta}+1)}{\sqrt{3 {\delta}^2+\sqrt{3} {\delta} (\theta^{\alpha} -\theta^{\beta})+(\theta^{\alpha})^2+\theta^{\alpha}\theta^{\beta}+\theta^{\beta}\,^2}},
\end{align*}\end{minipage}
}
\resizebox{1.05\linewidth}{!}{
  \begin{minipage}{\linewidth}
\begin{align*}
\nonumber& n_{m,m+2}&\\
\nonumber &=\frac{ 2 \left(3 \left(-3 {\delta}+2 \cos\theta^{\alpha}+\sqrt{3} \sin \theta^{\beta}+\cos \theta^{\beta}-3\right)^2+\left(3 {\delta}+2 \sqrt{3} \sin\theta^{\alpha}+\sqrt{3} \sin\theta^{\beta}-3 \cos\theta^{\beta}+3\right)^2\right)}{3 \sqrt{3} \text{a}^2 ({\delta}+1)^2}\\
&\approx \frac{8 \left(3 {\delta}^2+\sqrt{3} {\delta} (\theta^{\alpha}- \theta^{\beta})+\theta^2+\theta^{\alpha}  \theta^{\beta}+ (\theta^{\beta})^2\right)}{\sqrt{3} \text{a}^2 ({\delta}+1)^2},
\end{align*}\end{minipage}
}\\

Angles of the critical points (Divergence of $A$ and zero $n$):
\begin{equation*}
\theta^{\alpha}=-\theta^{\beta}=\tan ^{-1}\left[  \frac{\sqrt{3} \left(-{\delta}+\sqrt{1-3 {\delta} ({\delta}+2)}-1\right)}{3 {\delta}+\sqrt{1-3 {\delta} ({\delta}+2)}+3} \right]\approx -\sqrt{3} \delta.
\end{equation*}

 $\bigstar\;\mathbf{d}_{m,m+3}=\mathbf{b}^{\alpha}_{m}-\mathbf{b}^{\beta}_{m+3}=\frac{4\pi}{\sqrt{3}\text{a}}
\begin{pmatrix}
 \frac{\sin \theta^{\alpha}+\sin \theta^{\beta}}{1+\delta}  \\
 2-\frac{\cos \theta^{\alpha}+\cos \theta^{\beta}}{{\delta}+1}
\end{pmatrix},
 $

\begin{align}
\nonumber &A_{m,m+3}\\
\nonumber &=  \text{a} \frac{({\delta}+1)}{\sqrt{2}} \frac{1}{ \sqrt{-2 ({\delta}+1) \cos\theta^{\alpha}-2 ({\delta}+1) \cos  \theta^{\beta}+2 {\delta} ({\delta}+2)+\cos (\theta^{\alpha}- \theta^{\beta})+3}}
\end{align}
\begin{align}
\nonumber &n_{m,m+3}\\
\nonumber &=  -\frac{16 [ -2 ({\delta}+1) \cos\theta^{\alpha}-2 ({\delta}+1) \cos \theta^{\beta}+2 {\delta} ({\delta}+2)+\cos (\theta^{\alpha}- \theta^{\beta})+3]}{\sqrt{3} \text{a}^2 ({\delta}+1)^2},
\end{align}

with maximum of $A_{m,m+3}$ and minimum of $ n_{m,m+3}$ at $\theta^{\alpha}=\theta^{\beta}=0$.

 $\bigstar\;\mathbf{d}_{m,m+4}=\mathbf{b}^{\alpha}_{m}-\mathbf{b}^{\beta}_{m+4}=\frac{2\pi}{\text{a}(1+\delta)}
\begin{pmatrix}
 \frac{-3 {\delta}+2 \sqrt{3} \sin \theta+\sqrt{3} \sin \theta^{\prime}+3 \cos\theta^{\prime}-3}{3} \\
 \frac{3 {\delta}-2 \cos\theta+\sqrt{3} \sin \theta^{\prime}-\cos\theta^{\prime}+3}{\sqrt{3}}  
\end{pmatrix},
 $

\resizebox{1.05\linewidth}{!}{
  \begin{minipage}{\linewidth}
\begin{align*}
\nonumber &A_{m,m+4}\\
\nonumber &= \frac{2 \sqrt{3}  ({\delta}+1)\;\text{a}}{\sqrt{3 \left(3 {\delta}-2 \cos \theta^{\alpha}+\sqrt{3} \sin \theta^{\beta}-\cos  \theta^{\beta}+3\right)^2+\left(-3 {\delta}+2 \sqrt{3} \sin \theta^{\alpha}+\sqrt{3} \sin \theta^{\beta}+3 \cos \theta^{\beta}-3\right)^2}}\\
\approx &  \frac{\text{a}({\delta}+1)}{\sqrt{3 {\delta}^2+\sqrt{3} {\delta} (\theta^{\beta}-\theta^{\alpha})+(\theta^{\alpha})^2+\theta^{\alpha}\theta^{\beta}+\theta^{\beta}\,^2}},
\end{align*}\end{minipage}
}
\resizebox{1.05\linewidth}{!}{
  \begin{minipage}{\linewidth}
\begin{align*}
\nonumber& n_{m,m+4}\\
\nonumber &=\frac{2 \left(3 \left(3 {\delta}-2 \cos \theta^{\alpha}+\sqrt{3} \sin   \theta^{\beta}-\cos  \theta^{\beta}+3\right)^2+\left(-3 {\delta}+2 \sqrt{3} \sin  \theta^{\alpha}+\sqrt{3} \sin  \theta^{\beta}+3 \cos   \theta^{\beta}-3\right)^2\right)}{3 \sqrt{3} \text{a}^2 ({\delta}+1)^2}\\
&\approx \frac{8 \left(3 {\delta}^2+\sqrt{3} {\delta} ( \theta^{\beta}-\theta^{\alpha})+(\theta^{\alpha})^2+\theta^{\alpha}  \theta^{\beta}+ (\theta^{\beta})^2\right)}{\sqrt{3} \text{a}^2 ({\delta}+1)^2},
\end{align*}\end{minipage}
}\\

Angles of the critical points  (Divergence of $A$ and zero $n$):
\begin{equation*}
\theta^{\alpha}=-\theta^{\beta}=\tan ^{-1}\left[\frac{\sqrt{3} \left({\delta}-\sqrt{1-3 {\delta} ({\delta}+2)}+1\right)}{3 {\delta}+\sqrt{1-3 {\delta} ({\delta}+2)}+3}\right]\approx \sqrt{3} \delta.
\end{equation*}

 $\bigstar\;\mathbf{d}_{m,m+5}=\mathbf{b}^{\alpha}_{m}-\mathbf{b}^{\beta}_{m+5}=\frac{2\pi}{\text{a}(1+\delta)}
\begin{pmatrix}
 \frac{ 3 {\delta}-2 \sqrt{3} \sin\theta^{\alpha}+\sqrt{3} \sin\theta^{\beta}-3 \cos \theta^{\beta}+3}{3} \\
 \frac{{\delta}-2 \cos\theta^{\alpha}+\sqrt{3} \sin \theta^{\beta}+\cos\theta^{\beta}+1 }{\sqrt{3}}  
\end{pmatrix},
 $

\resizebox{1.05\linewidth}{!}{
  \begin{minipage}{\linewidth}
\begin{align*} 
&\nonumber A_{m,m+5}\\
=& \frac{2 \sqrt{3} \text{a} ({\delta}+1) }{\sqrt{3 \left({\delta}-2 \cos\theta^{\alpha}+\sqrt{3} \sin\theta^{\beta}+\cos \theta^{\beta}+1\right)^2+\left(3 {\delta}-2 \sqrt{3} \sin\theta^{\alpha}+\sqrt{3} \sin\theta^{\beta}-3 \cos \theta^{\beta}+3\right)^2}}  \\
\approx & \text{a} \frac{ {\delta}+1}{\sqrt{{\delta}^2+\sqrt{3} {\delta} ( \theta^{\beta}- \theta^{\alpha} )+(\theta^{\alpha})^2- \theta^{\alpha}   \theta^{\beta}+ (\theta^{\beta})^2}} ,
\end{align*}\end{minipage}
}
\resizebox{1.05\linewidth}{!}{
  \begin{minipage}{\linewidth}
\begin{align*}
&\nonumber n_{m,m+5}\\
&= \frac{2 \left(3 \left({\delta}-2 \cos \theta^{\alpha}+\sqrt{3} \sin  \theta^{\beta}+\cos  \theta^{\beta}+1\right)^2+\left(3 {\delta}-2 \sqrt{3} \sin\theta^{\alpha}+\sqrt{3} \sin \theta^{\beta}-3 \cos \theta^{\beta}+3\right)^2\right)}{3 \sqrt{3} \text{a}^2 ({\delta}+1)^2}\\
&\approx \frac{8 \left({\delta}^2+\sqrt{3} {\delta} (\theta^{\beta}-\theta^{\alpha})+(\theta^{\alpha})^2-\theta^{\alpha} \theta^{\beta}+\theta^{\beta}\,^2\right)}{\sqrt{3} \text{a}^2 ({\delta}+1)^2},
\end{align*}\end{minipage}
}\\

Angles of the critical points (Divergence of $A$ and zero $n$):
\begin{equation*}
\theta^{\alpha}=-\theta^{\beta}= \tan ^{-1}\left[\frac{-\sqrt{-({\delta}-1) ({\delta}+3)} {\delta}-\sqrt{-({\delta}-1) ({\delta}+3)}+\sqrt{3}}{{\delta} ({\delta}+2)-2}\right]   \approx  \frac{{\delta}}{\sqrt{3}}.
\end{equation*}

\clearpage

\includepdf[pages={1,2,3,4,5,6,7,8,9,10,11,12,13,14,15,16}]{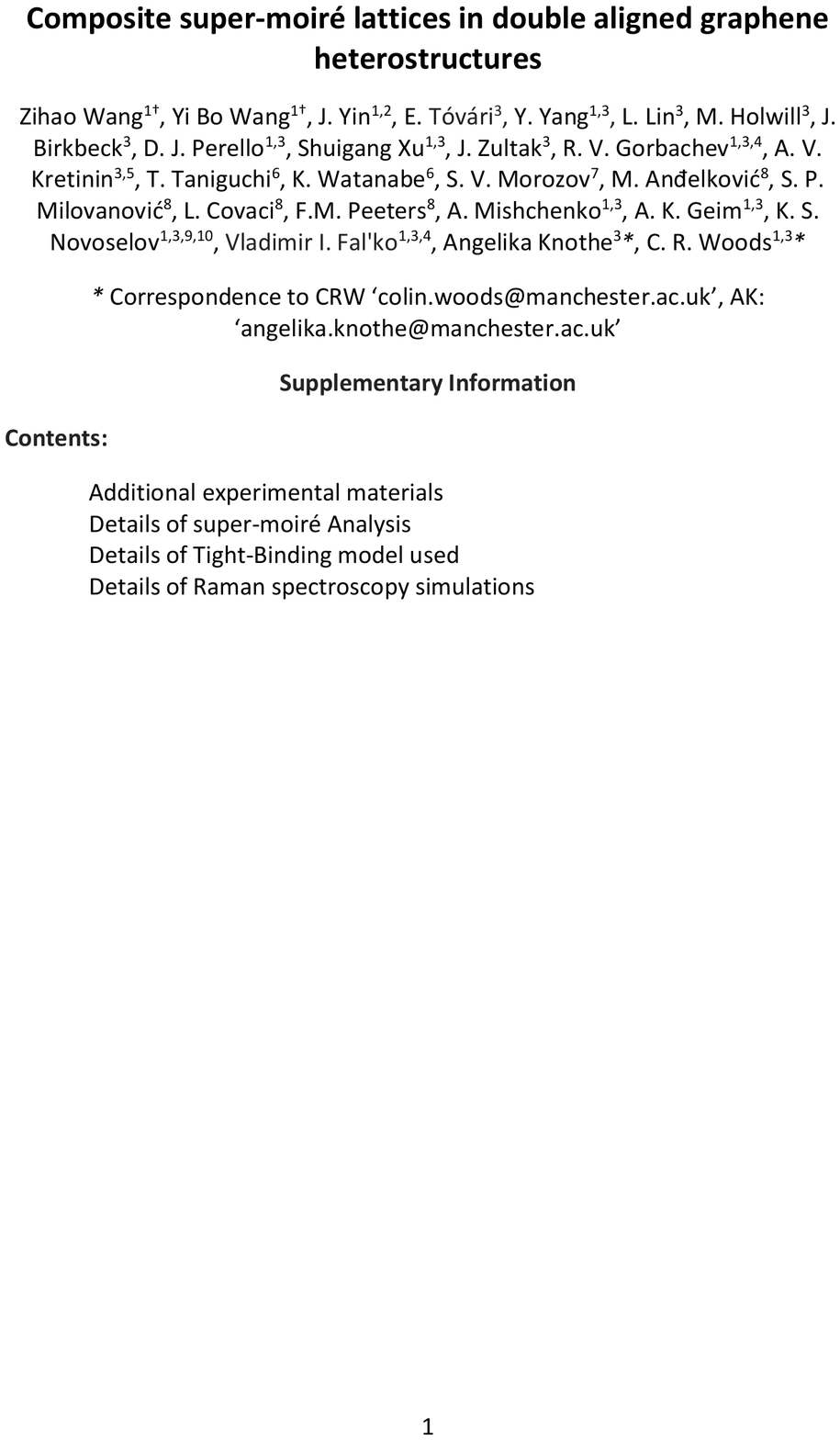}

\end{document}